%% file: cussima.tex
\documentclass[11pt,a4paper,twoside]{article}
\usepackage[centertags]{amsmath}
\usepackage{amsfonts}
\usepackage{amssymb}
\usepackage{multirow}
\usepackage{amsthm}
\usepackage{nicefrac}
\usepackage{pifont}
\usepackage[ansinew]{inputenc}
\usepackage{graphicx}
\usepackage[ruled]{algorithm}
\usepackage{algorithmic}

\if@shortnames
  \usepackage[authoryear,round]{natbib}
\else
  \usepackage[authoryear,round,longnamesfirst]{natbib}
\fi
\bibpunct{(}{)}{;}{a}{}{,}
\newcommand{\Address}[1]{\def\@Address{#1}}
\newcommand{\Plaintitle}[1]{\def\@Plaintitle{#1}}
\newcommand{\Shorttitle}[1]{\def\@Shorttitle{#1}}
\newcommand{\Plainauthor}[1]{\def\@Plainauthor{#1}}
\newcommand{\URL}[1]{\def\@URL{#1}}

\newlength{\footerskip}
\RequirePackage{hyperref}

\let\code=\texttt
\let\proglang=\textsf

\newcommand{\email}[1]{\href{mailto:#1}{\normalfont\texttt{#1}}}
\newcommand{\doi}[1]{\href{http://dx.doi.org/#1}{\normalfont\texttt{doi:#1}}}

\author{Ralph Brinks\\German Diabetes Center}
\title{Simulation of Populations in a Time-, Age- and Duration Dependent Illness-Death Model}

\Plainauthor{Ralph Brinks} 

\Address{
  Ralph Brinks\\
  Institute for Biometry \& Epidemiology\\
  German Diabetes Center\\
  D-40225 D\"usseldorf, Germany\\
  E-mail: \email{ralph.brinks@ddz.uni-duesseldorf.de}\\
}

\newcommand{\makefooter}{%
  \vspace{\footerskip}

    \begin{samepage}
    \textbf{\large \nopagebreak}\\[.3\baselineskip] \nopagebreak
    \@Address \nopagebreak
    \end{samepage}
}

\AtEndDocument{\makefooter}
\AtBeginDocument{\maketitle}

\begin{document}
\begin{abstract}
Relevant events in a three state illness-death model (IDM) of a
chronic disease are the diagnosis of the disease and death with or
without the disease. In this article a simulation framework for
populations moving in the IDM is presented. The simulation is
closely related to the concept of Lexis diagrams in event history
analysis. Details of the implementation and an example of a
hypothetical disease are described.
\end{abstract}


\input{cussima_01}

\bibliography{cussima}

\end{document}

%% file: cussima_01.tex
\section{A simple illness-death model}
\begin{figure*}[ht]
\centerline{\includegraphics[keepaspectratio,
width=14cm]{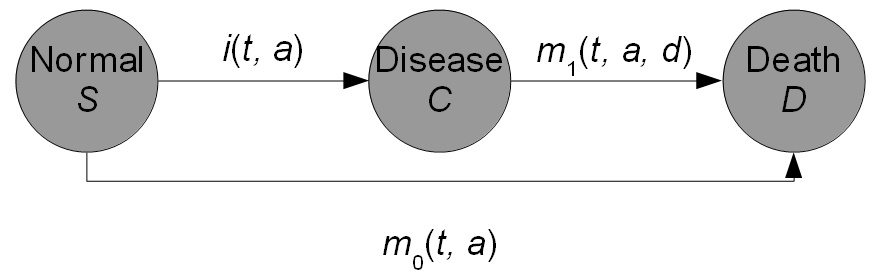}} \caption{Three states model of normal,
diseased and dead subjects. Transition densities may depend on
calender time $t,$ age $a,$ and in case of $m_1$ also on the
duration $d$ of the disease.} \label{fig:3states}
\end{figure*}

A popular framework for studying irreversible diseases is the
illness-death model (IDM) consisting of three states as depicted
in Figure~\ref{fig:3states}: \emph{Normal, Disease} and
\emph{Death} \citep{Kei91,Kal02,Aal08}. The associated transition
rates\footnote{synonymously: densities (in units ``per
person-year'', not to be confused with risks or probabilities
\citep{Van12}).} are denoted with the symbols as in
Figure~\ref{fig:3states}: incidence $i$, and mortality rates $m_0$
and $m_1$. In general, the rates depend on different time scales:
calendar time $t$, age $a$ and in case of $m_1$ on the duration
$d$ of the disease.

In this article a framework for simulating populations moving in
the IDM is presented. For each person $j, ~j = 1, \dots, n$ in the
population, the relevant events \emph{diagnosis} and \emph{death}
are simulated. This is accomplished in two steps:

\begin{enumerate}
\item Contracting the disease and dying without the disease is
modelled as competing risk. Given the time $t_0^{(j)}$ of birth of
person $j$, the cumulative distribution function $F_1^{(j)}$ of
the ``first failure time'' $T_1^{(j)}$ is
\begin{equation}\label{e:distrFunc1}
    F_1^{(j)}\left (t \right ) = 1 - \exp \left ( -
    \int_0^t i \left (t^{(j)}_0 + \tau, ~\tau \right ) + m_0 \left (t^{(j)}_0 + \tau, ~\tau \right )
    \mathrm{d}\tau   \right ).
\end{equation}
The term ``first failure time'' $T_1^{(j)}$ refers to the time of
diagnosis or death without disease. $T^{(j)}_1$ is measured in
time units after birth of person $j$. Thus, $T_1^{(j)}$ is the age
when the first transition from the state \emph{Normal} occurs.
Assumed we know that for person $j$ at $T_1^{(j)}$ a transition
occurs, then the odds of transiting into state \emph{Disease}
versus transiting into state \emph{Death} is
\begin{equation}\label{e:Odds}
    \frac{i \left (t^{(j)}_0 + T^{(j)}_1, T^{(j)}_1 \right )}{m_0 \left (t^{(j)}_0 +
T^{(j)}_1, T^{(j)}_1 \right )}.
\end{equation}

\item If the event at $T_1^{(j)}$ is the death (without the
disease), the simulation for person $j$ is finished. If, however,
the event is the diagnosis of the disease, the ``second failure
time'' $T_2^{(j)}$ to death (with disease) has the distribution
function $F_2^{(j)}:$
\begin{equation}\label{e:distrFunc2}
    F_2^{(j)} \left ( t~\vert ~T_1^{(j)} \right ) = 1 - \exp \left (-
    \int_0^t m_1 \left (t^{(j)}_0 + T_1^{(j)} + \tau, ~T_1^{(j)} + \tau, ~\tau \right )
    \mathrm{d}\tau   \right ).
\end{equation}
\end{enumerate}

The next section describes in detail how the integrals in Eqs.
\eqref{e:distrFunc1} and \eqref{e:distrFunc2} are calculated in
the implementation of the simulation. After calculating the
integrals, the question arises how the times $T_1$ and $T_2$ can
be obtained from $F_1$ and $F_2$. This is done by the
\emph{inverse transform sampling method}: Let $F$ be a cumulative
distribution function and $u \in (0, 1)$. For $F^{-1}(u) :=
\inf\;\{x \mid F(x)\geq u\},$ it holds: If $U$ is a uniform random
variable on $(0, 1)$, then $F^{-1}(U)$ follows the distribution
$F$. Thus, the simulation of $T_1$ and $T_2$ is easy, if a random
number generator for $U$ such as \code{runif} in \proglang{R} is
available.

\bigskip

For each of the $n$ persons in the population we store four pieces
of data:
\begin{enumerate}
    \item a unique identifier $j,$
    \item the date $t_0^{(j)}$ of birth (\verb"dob") of person $j,$
    \item the age at diagnosis (\verb"adi") of person $j,$ and
    \item the age of death (\verb"ade") of person $j.$
\end{enumerate}
If the person $j$ does not contract the disease, the age at
diagnosis \verb"adi" is set to \verb"NA" (missing).

\bigskip

In summary, we get the following
\begin{algorithm}[h!]
\caption{Simulation of populations moving in the
IDM}\label{alg:spi}
\begin{algorithmic}[1]
\FOR{$j=1$ \TO $n$}
   \STATE \verb"dob" $\leftarrow t_0^{(j)}$
   \STATE calculate event time $T_1^{(j)}$ according to Eq. \eqref{e:distrFunc1}
   \STATE simulate type of event that has happened at $T_1^{(j)}$ by Eq. \eqref{e:Odds}
   \IF{event is diagnosis}
      \STATE \verb"adi" $\leftarrow T_1^{(j)}$
      \STATE calculate time $T_2^{(j)}$ of death using Eq. \eqref{e:distrFunc2}
      \STATE \verb"ade" $\leftarrow T_1^{(j)} + T_2^{(j)}$
   \ELSE
      \STATE \verb"adi" $\leftarrow$ \verb"NA"
      \STATE \verb"ade" $\leftarrow T_1^{(j)}$
   \ENDIF
   \STATE write \verb"j, dob, adi, ade" to file
\ENDFOR
\end{algorithmic}
\end{algorithm}

This allows us to the define the file format for storing the
results of Algorithm \ref{alg:spi}. For each person $j$ the four
entries \verb"j, dob, adi, ade" stored in a row of an ASCII text
file. Delimiter between the four entries are semicolons (;),
decimal separators are dots (.). The file extension is \verb"spd"
(\textbf{\emph{s}}imulation of \textbf{\emph{p}}opulations moving
in the illnees-\textbf{\emph{d}}eath model).

\bigskip

A few words are devoted to the applications and benefits, such a
simulation has. The motivation for the algorithm comes from
analytical epidemiology where relations between common
epidemiological measures are studied. Examples for those measures
are the prevalence, the duration of a disease, the age of onset or
diagnosis, and lost life years (due to the disease). Obviously,
the characteristics of the age of diagnosis can be obtained
directly from the simulation. A typical question in that respect
may be: what is the mean age of diagnosis of those subjects born
between $t_s$ and $t_e$? Another interesting aim is the estimation
of the incidence rate $i$ from cross-sectional information. At a
specific point in time $t'$, each of the subjects $j = 1, \dots,
n,$ has a unique ``status''. Neglecting those who are unborn or
dead at $t'$, the status is either \emph{normal} (non-diseased) or
\emph{diseased}. Thus, the status can be seen as a binary random
variable, and data of this kind is typically called \emph{current
status data}. The current status is closely linked with the
incidence $i$ and the mortalities $m_0$ and $m_1$. Estimating the
incidence from current status data, for example, has been a topic
in research for decades \citep{Hen10}. The framework presented
here may be useful in this field.

\section{Line integrals in the Lexis diagram}
\begin{figure}[th]
\begin{center}
\includegraphics[width=9.6cm, keepaspectratio]{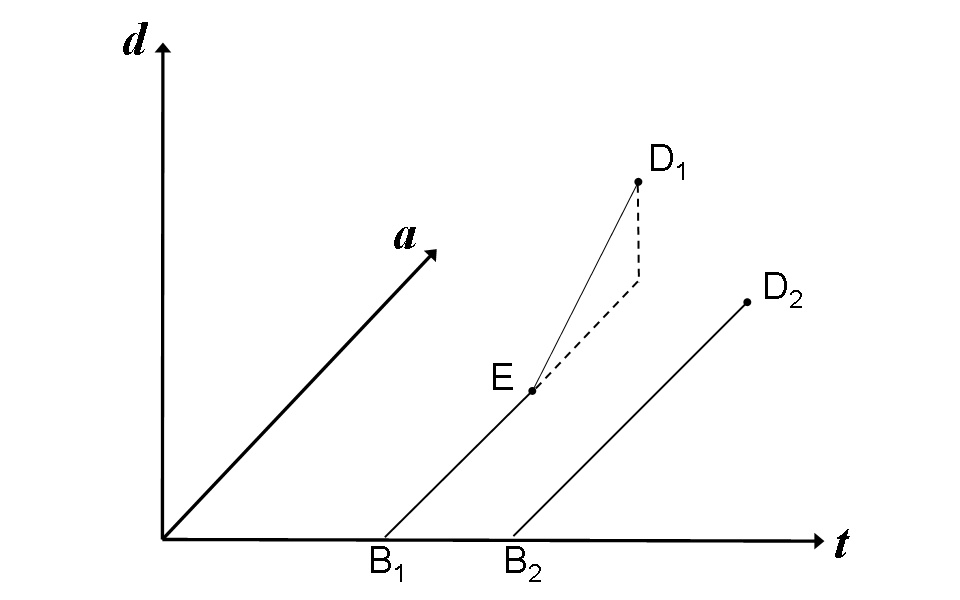}
\end{center}
\caption{Three-dimensional Lexis diagram with two life lines.
Abscissa, ordinate and applicate represent calendar time $t$, age
$a$ and duration $d$, respectively. The life lines start at birth
$B_\nu$ and end at death $D_\nu, ~\nu=1,2.$ The first subject gets
the disease at $E$. Then, the life line changes its direction. The
second subject does not get the disease, the corresponding life
line remains in the $t$-$a$-plane. }\label{fig:Lexis}
\end{figure}

In this section the calculation of the integrals in Eqs.
\eqref{e:distrFunc1} and \eqref{e:distrFunc2} are described. Since
we are interested in integrating arbitrary integrands $i, m_0,$
and $m_1$, we use numerical integration. We assume that the
integrands are given by numerical values on a regular grid. In
event history analysis \citep{Kei06}, a useful concept is the
\emph{Lexis diagram}, which is a co-ordinate system with axes
calendar time $t$ (abscissa) and age $a$ (ordinate). The calendar
time dimension sometimes is referred to as period. Each subject is
represented by a line segment from time and age at entry to time
and age at exit. Entry and exit may be birth and death,
respectively, or entry and exit in a epidemiological study or
clinical trial. There are excellent and extensive introductions
about the theory of Lexis diagrams (see for example
\citep{Kei90,Kei91,Car06} and references therein), which allows to
be short here. When it comes to irreversible diseases, the
commonly used two-dimensional Lexis diagram with axes in time and
age direction may be generalized to a three-dimensional
co-ordinate system with disease duration $d$ represented by the
applicate (z-axis). If a subject does not get the disease during
life time, the life line remains in the time-age-plane parallel to
the line bisecting abscissa and ordinate. With other words, the
life line for the time without disease is parallel to $e_1 :=
(1,1,0)^t$ (where the triple $(t,a,d)$ denotes the co-ordinates in
time, age and duration direction, respectively). However if at a
certain point in time $E$ the disease is diagnosed, the life line
changes its direction, henceforth parallel to $e_2 := (1,1,1)^t$.
The situation is illustrated in Figure~\ref{fig:Lexis}. The life
lines of two subjects are shown in the three-dimensional Lexis
space. At time of birth (denoted $B_\nu, ~\nu=1,2$) both subjects
are disease-free; both life lines are parallel to $e_1.$ The first
subject gets the disease at $E$, and henceforth the life line is
parallel to $e_2$ until death at $D_1$. The second subject remains
without the disease for the whole life, which ends at $D_2$.

\bigskip

Having the concept of the Lexis diagram at hand, we observe that
$F_1$ and $F_2$ in Eqs. \eqref{e:distrFunc1} and
\eqref{e:distrFunc2} are line integrals in the Lexis space. We
start with calculating the first failure times $T_1.$ For subject
$j$ the associated life line starts at $(t, a) = (t_0^{(j)}, 0)$.
We chose an age $\omega > 0,$ when it is sure that a transition to
one of the states \emph{Disease} or \emph{Death} has occurred, say
$\omega = 150$ (years). For calculating $F_1^{(j)},$ we trace the
hypothetical life line from $B_j := (t_0^{(j)}, 0)$ to $D_j :=
(t_0^{(j)} + \omega, \omega).$ Thus, the hypothetical life line
has a representation
\begin{equation*}
   {\cal L}_j: ~B_j + \alpha \cdot (D_j - B_j), ~\alpha \in [0, 1].
\end{equation*}

As described in \citep{Bri12} following the life line is related
to raytracing in the field of computer graphics, where efficient
algorithms for this purpose exist. In Siddon's algorithm
\citep{Sid85}, the key idea is to follow ${\cal L}_j$ by
calculating intersections with volume elements (voxels), which
form a regular partition of the Lexis space. Let
$$A^\star_j = \{\alpha^{(j)}(p) ~\vert ~p = 1, \dots, P^{(j)} \}$$
with $0 = \alpha^{(j)}(1) < \dots < \alpha^{(j)}(P^{(j)}) = 1$ be
a parametrization of the points where ${\cal L}_j$ intersects the
voxel faces plus the start and and end points $B_j$ and $D_j$.
Details for the calculation of $A^\star_j$ are described in
\citep{Bri12}. The parametrization $A^\star_j$ is ideally suited
for approximating the integral in Eq. \eqref{e:distrFunc1} by the
trapezoidal rule \citep{Dah74}. The reason lies in the fact that
in calculating $F_1^{(j)}(\omega)$ the values $F_1^{(j)}\left
(t_0^{(j)} + \alpha^{(j)}(p) \, \omega \right ), ~p=1,\dots,
P^{(j)},$ are a byproduct. Algorithm \ref{alg:F1} shows the
necessary steps.

\begin{algorithm}[h!]
\caption{Calculating $F_1$}\label{alg:F1}
\begin{algorithmic}[1]
\FOR{$j=1$ \TO $n$}
   \STATE Calculate $A^\star_j = \{\alpha^{(j)}(p) ~\vert ~p = 1, \dots, P^{(j)}
   \}$.
   \STATE $\ell_1 \leftarrow 0$
   \STATE $\tau_1 \leftarrow 0$
   \STATE $f_1 \leftarrow i\left (t^{(j)}_0, ~0 \right ) + m_0\left (t^{(j)}_0, ~0 \right )$
   \STATE $F_1^{(j)}(\tau_1) \leftarrow 0$
   \FOR{$p=2$ \TO $P^{(j)}$}
      \STATE $\tau_p \leftarrow \alpha^{(j)}(p) \cdot \omega$
      \STATE $f_p \leftarrow i\left (t^{(j)}_0 + \tau_p, ~\tau_p \right) + m_0\left (t^{(j)}_0 + \tau_p, ~\tau_p \right)$
      \STATE $\ell_p \leftarrow \ell_{p-1} + \tfrac{1}{2} \cdot (\tau_p - \tau_{p-1}) \cdot (f_p + f_{p-1})$
      \STATE $F_1^{(j)}\left (t^{(j)}_0 + \tau_p \right ) \leftarrow 1 - \exp(-\ell_p)$
   \ENDFOR
\ENDFOR
\end{algorithmic}
\end{algorithm}

Since the values of $i$ and $m_0$ are given on the voxel grid
only, the calculation of $f_p, ~p=1, \dots, P^{(j)},$ needs
bilinear interpolation of the values of the adjacent voxels
\citep{Pre88}.

After preparing $F_1^{(j)}, j=1, \dots, n,$ the times $T_1^{(j)}$
can be calculated by the inverse transform sampling method. Since
we have $F_1^{(j)}$ calculated at points $\zeta_p := t^{(j)}_0 +
\tau_p, ~p=1, \dots, P^{(j)},$ the inverse transform sampling
would yield only those $\zeta_p$. A better accuracy can be
obtained by interpolating $F_1^{(j)}$ affine-linearly between
consecutive $\zeta_p.$ For $t \in (\zeta_{p-1}, ~\zeta_p), ~p = 2,
\dots, P^{(j)},$ let $\xi := \tfrac{t-\zeta_{p-1}}{\zeta_{p} -
\zeta_{p-1}}.$ Then, it holds

$$F_1^{(j)}(t) \approx \left (1 - \xi \right ) \cdot F_1^{(j)}\left (\zeta_{p-1} \right ) +
                     \xi \cdot F_1^{(j)} \left (\zeta_{p} \right ).$$

For those subjects $j'$ who contract the disease, the associated
$F_2^{(j')}( \cdot \mid T_1^{(j')})$ can be derived in a similar
way as in Algorithm \ref{alg:F1}. The associated line segment
starts at $(t, a, d) = (t_0^{(j')} + T_1^{(j')}, ~T_1^{(j')}, 0
)$. Again, a hypothetical maximal disease duration $\omega'$ is
assumed, say $\omega' = 50$ (years), such that the line segment
ends at $(t, a, d) = (t_0^{(j')} + T_1^{(j')} + \omega',
~T_1^{(j')} + \omega', ~\omega')$. Thus, the line segment is
parallel to $e_2 = (1, 1, 1)^t.$ The Siddon algorithm computes the
corresponding set of intersections with the voxel grid
accordingly. The ages $T_2^{(j')}$ of death with disease are
obtained from Algorithm \ref{alg:F1} mutatis mutandis. The
interpolation of $m_1$ needs to be trilinear.

\section{Example}
\begin{figure*}[ht]
\centerline{\includegraphics[keepaspectratio,
width=14cm]{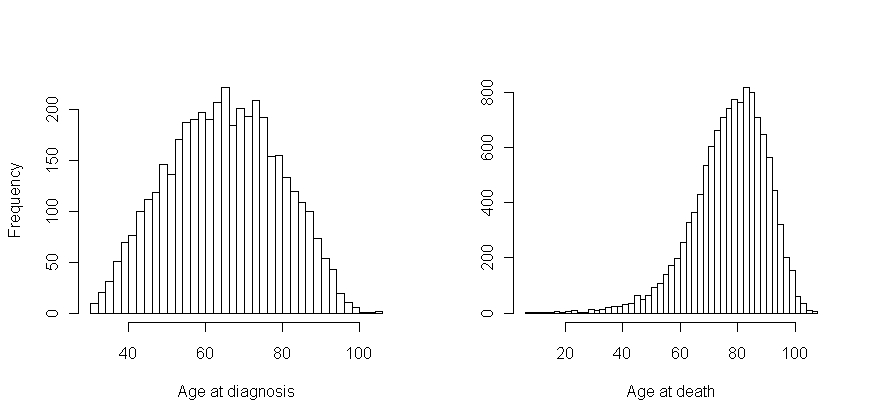}} \caption{Histograms of the age at onset
(left) and age at death (right) in the simulation.}
\label{fig:Hists12k}
\end{figure*}

This section presents the results of a simulation. In each of
sixty consecutive years $t = 0, \dots, 59,$ two hundred persons
are born and followed from birth to death. The incidence of a
hypothetical chronic disease is assumed to be $i(t, a) = \tfrac{
(a-30)_+}{3000}$, the mortality of the non-diseased is $m_0(t,a) =
\exp(-10.7 + 0.1 a + t \ln(0.998))$ and the mortality of the
diseased is $m_1(t, a, d) = m_0(t, a) \cdot ( 0.04 (d - 5)^2 + 1
).$ In total, 4184 of the 12000 simulated persons contract the
disease. The simulated data easily allows derivation of important
epidemiological measures. For example, the histograms of the age
at onset and age at death are shown in Figure \ref{fig:Hists12k}.

The median age at death of those who contracted the disease is
77.3 (years) whereas the median age at death of those without the
disease is 79.6 (years). The mean duration of the disease in the
4184 ill subjects is 12.5 (years).

\bigskip

\begin{figure*}[h!]
\centerline{\includegraphics[keepaspectratio,
width=9.6cm]{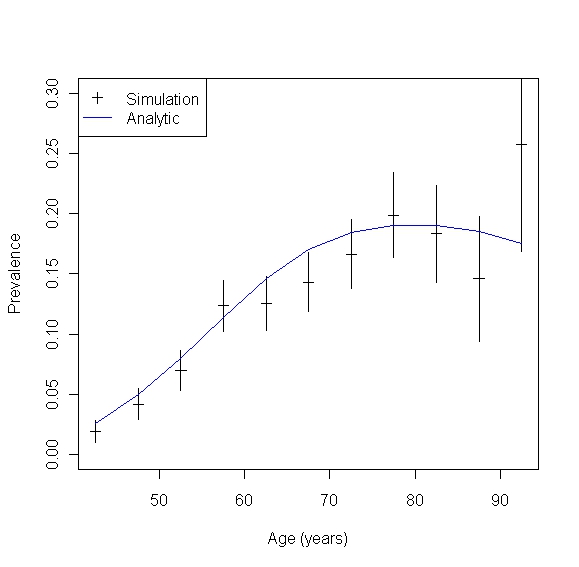}} \caption{Simulated (black) and
analytically calculated (blue) prevalence of a hypothetical
disease. The simulated prevalence is depicted with 95\% confidence
intervals.} \label{fig:Test12k}
\end{figure*}

To cross-check the results of the simulation, we compare them to a
theoretical calculation. In year $t = 100,$ exactly 7368 persons
are alive, 799 having the hypothetical disease. Figure
\ref{fig:Test12k} shows the age-specific prevalence in the year
100. The black lines indicate the prevalence of several age groups
together with 95\% confidence bounds as given by the simulation.
The blue line represents the prevalence calculated analytically by
the exact formula in \cite[Section 7.2]{Kei91}. The results agree
quite well within the confidence bounds.

\section{Summary}
This article is about simulating populations in an illness-death
model consisting of the three states \emph{Normal, Disease,} and
\emph{Death}. The disease is assumed to be irreversible. After
birth of an subject in the population, two cases may occur:
\begin{enumerate}
 \item the subject dies without the disease, or
 \item the subject contracts the disease and dies with the disease.
\end{enumerate}
In the first case, the life lines of the Lexis diagram are solely
located in the $t$-$a$-plane parallel to $e_1 = (1,1,0)^t$. In the
second case, a part of the life lines is parallel to the $e_2 =
(1,1,1)^t$ direction. Changing the direction of the life line
allows modelling the covariable \emph{duration of the disease.} In
many diseases, the duration plays an important role for the
mortality. Examples are diseases related to arteriosclerosis such
as diabetes \citep{Car08} or lupus erythematosus \citep{Ber06}.

The simulation is based on raytracing techniques and provides a
fast way to follow the individual life lines of subjects in the
Lexis diagram. Computation time is an issue, because the number of
subjects may be large (several thousands). The simulation
calculates event times (diagnosis of the disease or death) and
uses inverse transform sampling via cumulative distribution
functions. The integrals occurring in the distribution functions
are approximated by the trapezoidal rule of numerical integration,
which ideally fits to the raytracing technique.